\documentclass[lnbip]{svmultln}
\usepackage{makeidx}  
\usepackage{textcomp}
\usepackage{graphics}
\usepackage{graphicx}

\begin{document}
\mainmatter              
\title{Agility in Software 2.0 -- Notebook Interfaces and MLOps with Buttresses and Rebars}
\titlerunning{Agility in Software 2.0}  
%
\author{Markus Borg}
\authorrunning{Markus Borg}   
%
%
\institute{RISE Research Institutes of Sweden, Lund, Sweden,\\
Dept. of Computer Science, Lund University, Lund, Sweden\\
markus.borg@ri.se}

\maketitle              

\begin{abstract}        
Artificial intelligence through machine learning is increasingly used in the digital society. Solutions based on machine learning bring both great opportunities, thus coined ``Software 2.0,'' but also great challenges for the engineering community to tackle. Due to the experimental approach used by data scientists when developing machine learning models, agility is an essential characteristic. In this keynote address, we discuss two contemporary development phenomena that are fundamental in machine learning development, i.e., notebook interfaces and MLOps. First, we present a solution that can remedy some of the intrinsic weaknesses of working in notebooks by supporting easy transitions to integrated development environments. Second, we propose reinforced engineering of AI systems by introducing metaphorical buttresses and rebars in the MLOps context. Machine learning-based solutions are dynamic in nature, and we argue that reinforced continuous engineering is required to quality assure the trustworthy AI systems of tomorrow. 
\end{abstract}

\section{Introduction}
No one has missed the AI surge in the last decade. There is an ever-increasing number of AI applications available as enterprises across domains seek to harness the promises of AI technology. Enabled by the growing availability of data, most of the AI success stories in recent years originate in solutions dominated by Machine Learning (ML)~\cite{giray2021software}. Where human programmers previously had to express all logic in source code, ML models can now be trained on huge sets of annotated data -- for certain tasks, this works tremendously well. Andrej Karpathy, AI Director at Tesla, somewhat cheekily refers to development according to the ML paradigm as ``Software 2.0''\footnote{bit.ly/3dKeUEH}. For many applications seeking mapping from input to output, it is easier to collect and annotate high-quality data than to program a mapping function in code explicitly.

Agile software development has become the norm in the software engineering industry. Flexibly adapting to change has proven to be a recipe to ripe some of the benefits of software -- significant changes can often occur at any time, both during a development project and post-release. Quickly adapting to shifting customer needs and technology changes is often vital to survival in a competitive market. In this light, the concept of DevOps has emerged as an approach to minimize time to market while maintaining quality~\cite{Ebert2016}. While agile development is particularly suitable for customer-oriented development in the Internet era, it is also increasingly used in embedded systems development of more critical nature~\cite{diebold2018agile} with adaptations such as SafeScrum~\cite{hanssen2018safescrum}. Moreover, while agile software development is flexible, we argue that ML development iterates even faster -- and thus necessitates ``agility on steroids.''

Data scientists often conduct the highly iterative development of ML models. Data scientists, representing a new type of software professionals, often do not have the software engineering training of conventional software developers~\cite{kim2016emerging}. This observation is analogous to what has been reported for developers of scientific computing in the past, e.g., regarding their familiarity with agile practices~\cite{sletholt2011we}. Instead of prioritizing the crafts of software engineering and computer science, many data scientists focus on mastering the art of taming data into shapes that are suitable for model training -- typically using domain knowledge to hunt quantitative accuracy targets for a specific application. The ML development process is experimental in nature and involves iterating between several intertwined activities, e.g., data collection, data preprocessing, feature engineering, model selection, model evaluation, and hyperparameter tuning. An unfortunate characteristic of ML development is that nothing can be considered in isolation. A foundational ML paper by Google researchers described this as the CACE principle ``Changing Anything Changes Everything''~\cite{sculley_hidden_2015}. When developing ML models in Software 2.0, no data science activities are ever independent. 

In this keynote address, we will discuss two phenomena that have emerged to meet the characteristics of ML development. First, \textbf{Notebook interfaces} to meet the data scientists' needs to move swiftly. Unfortunately, the step from prototyping in Notebook interfaces to a mature ML solution is often considerable -- and cumbersome for many data scientists. In Section~\ref{sec:notebooks}, we will present a solution by Jakobsson and Henriksson that bridges the gap between the data scientists' preferred notebook interfaces and standard development in Integrated Development Environments (IDE). Second, analogous to DevOps in conventional agile software development, in Section~\ref{sec:mlops}, we will look at how \textbf{MLOps} has emerged to close the gap between ML development and ML operations. More than just an agility concept, we claim that it is required to meet the expectations on the trustworthy AI of the future -- illustrated in the light of the recently proposed Artificial Intelligence Act in the European Union. We refer to our concept of reinforcing the development and operations of AI systems, afflicted by the CACE principle, using two metaphors from construction engineering: buttresses and rebars.

\section{Connecting Notebook Interfaces and IDEs} \label{sec:notebooks}
Many data scientists are not trained software engineers and thus might not be fully aware of available best practices related to various software engineering activities~\cite{kim2016emerging}. Moreover, even with awareness of software engineering best practices, data science introduces new challenges throughout the engineering lifecycle~\cite{amershi2019software,wan2019does} -- from requirements engineering~\cite{vogelsang_requirements_2019} to operations~\cite{sculley_hidden_2015}. Due to the intrinsically experimental nature of data science, practitioners seek development environments that allow maximum agility, i.e., high-speed development iterations. 

The go-to solution for many data scientists is to work iteratively in cloud-based notebook interfaces. While this allows rapid experimentation, it does not easily allow the application of the various tools available in a modern IDE~\cite{notebookPainPoints}. The first part of this keynote address presents a solution developed as part of a MSc thesis project by Jakobsson and Henriksson at Backtick Technologies~\cite{backtick} that enables data scientists to easily move between notebook interfaces and an IDE thanks to a networked file system. The idea is to let data scientists work in their favorite editor and use all the tools available for local development while still being able to use the cloud-based notebook interface for data exploration -- and reaping its benefits of easy access to distributed cloud computing. Jakobsson and Henriksson integrated and evaluated the solution as part of Cowait Notebooks, an experimental cloud notebook solution developed by Backtick Technologies. Cowait\footnote{https://cowait.io} is an open-source framework for creating containerized distributed applications with asynchronous Python.

\subsection{Agility Supported by Notebook Interfaces}
A substantial part of today's data science revolves around notebook interfaces, also known as computational notebooks. Notebook interfaces are typically cloud-based and consist of environments with interactive code interpreters accessible from web browsers that allow raöid, iterative development. The notebooks themselves usually run on a remote machine or a computer cluster, allowing the user easy access to compute resources available in data centers. While the notebook interfaces gradually mature, i.e., more features become available, the environments are still far from as capable as the IDEs software developers run locally. Consequently, the support for version control software, static analysis, linting, and other widely used development tools is limited in notebook interfaces~\cite{notebookPainPoints}. 

The implementation of a notebook interface differs from a conventional IDE. A notebook runs an interpreter in the background that preserves the state for the duration of a programming session. A user observes a notebook as a sequence of cells that are either textual (allowing data scientists to document the process) or containing code. These two different types of cells are interwoven in the notebook. Notebook interfaces usually excel at presenting plots and tables that support data exploration. A code cell contains one or more statements and can be executed independently from any other code cell. Users can execute code cells in any order, but the cells all mutate the shared state of the background interpreter. This freedom of execution order greatly supports the agility of data science as users can re-run portions of a program while keeping other parts of the previously generated state. While this enables fast iterations toward a useful solution, it also makes it difficult to trace the path of execution that led to a specific result. Even worse, subsequent executions of the notebook may yield different results.

The concept of computational notebooks was envisioned by Knuth already in 1984~\cite{notebookKnuth}. Knuth proposed the \textit{literate programming} paradigm and showed how the idea could support program comprehension by mixing snippets of source code and natural language explanations of its embedded logic. As elaborated in Knuth's seminal book on the topic~\cite{knuth_book}, the key point is that literate programming explicitly shifts who is the most important reader of the programming artifact. In literate programming, source code is primarily written for \textit{humans} instead of computers -- and the artifact can be seen as a piece of literature. Many developers of scientific computing follow this paradigm to develop maintainable software artifacts~\cite{hannay2009scientists}. 

A more general version of literate programming is \emph{literate computing}, where the source code cells and natural language explanations are accompanied by visual content such as tables, graphs, and images. Today's widely used notebook interfaces, such as the popular Jupyter Notebook\footnote{https://jupyter.org} and Databrick's Collaborative Notebook\footnote{https://databricks.com/product/collaborative-notebooks}, are examples of literate computing. For a recent overview of the notebook landscape, we refer the curious reader to an article by Vognstrup Fog and Nylandsted Klokmose~\cite{notebookLandscape}. Their summary presents both a historical perspective and a discussion of design decisions for future notebook interfaces.

Notebook interfaces have certainly evolved substantially since Knuth first envisioned them. However, there are still certain impediments for data scientists working in notebooks. Chattopadhyay \textit{et al.} analyzed contemporary issues with notebook interfaces and reported nine pain points~\cite{notebookPainPoints}. According to the authors, the most pressing pain points for developers of notebook interfaces to tackle are 1) code refactoring, 2) deployment to production, 3) exploring notebook history, and 4) managing long-running tasks. Notebook interfaces constitute a highly active research topic, and researchers have proposed several solutions to address their limitations~\cite{notebookForaging,notebookUntangle,notebookManageMesses}. However, while notebook interfaces are a prominent medium for software development, there is still a substantial need for research and development~\cite{notebookNotes}.

This talk will introduce a solution proposal by Jakobsson and Henriksson that bridges the benefits of notebook interfaces and local IDEs. Lau \textit{et al.} examined 60 different notebook interfaces and categorized them according to 10 dimensions of analysis: 1) data sources, 2) editor style, 3) programming language, 4) versioning, 5) collaboration, 6) execution order, 7) execution liveness, 8) execution environment, 9) cell outputs, and 10) notebook outputs. In the MSc thesis project by Jakobsson and Henriksson, the authors focused on the dimensions of \emph{execution environment} and \emph{data sources} for Cowait Notebooks. Their solution allows Cowait Notebooks to execute code in a remote multi-process execution environment using local files as data sources. This solution contrasts with Jupyter Notebook for which both code execution and data is local. The solution is also different from Databrick's Collaborative Notebook, where code is executed in a remote multi-process execution environment, but the data sources cannot be local. In the next section, we present the open-source Cowait framework. 

\subsection{Cowait -- A Framework for Simplified Container Orchestration}
Cowait is a framework that simplifies the execution of Python code on the container orchestration system Kubernetes. The two main constituents of Cowait are 1) a workflow engine built on top of Docker and Kubernetes and 2) a build system to easily package source code into containers. Together, the workflow engine and the build system form an abstraction of containers and container hosts that helps developers leverage the power of containerization through Docker and cluster deployment using Kubernetes without knowing all technical details. Backtick Technologies designed Cowait to hide the intrinsic complexity of Docker and Kubernetes behind simple concepts that are familiar to general software developers. Cowait is developed under an Apache License and the source code is available on GitHub\footnote{https://github.com/backtick-se/cowait}.

Cowait provides four key features with a focus on user-friendliness, i.e., Cowait\ldots
\begin{enumerate}
    \item \ldots helps the development of distributed workflows on your local machine with minimal setup.
    \item \ldots simplifies dependency management for Python projects.
    \item \ldots allows developers to unit test their workflow tasks.
    \item \ldots lowers the bar for users to deploy solutions on Kubernetes clusters.
\end{enumerate}

In line with other workflow engines, Cowait organizes code into \textit{tasks}. A task is essentially a function that can accept input arguments and return values. As for functions in general, a task can invoke other tasks –- with one key difference: a call to invoke another task will be intercepted by the Cowait runtime environment and subsequently executed in a separate container. Cowait can also direct the execution of this separate container to a particular machine. The fundamental differentiator offered by Cowait is that tasks can interface directly with the underlying cluster orchestrator. In practice, this means that tasks can start other tasks without going through a central scheduler service. Instead, tasks create other tasks on demand, and they communicate with their parent tasks using web sockets. Further details are available in the Cowait Documentation\footnote{https://cowait.io/docs/}.

The task management system in Cowait relies on containers and thus supports the execution of arbitrary software. Thanks to this flexibility, Cowait can execute notebook interfaces. In their MSc thesis project, Jakobsson and Henriksson demonstrate the execution of the open-source JupyterLab notebook interface in a Cowait solution -- we refer to this as running a Cowait Notebook. JupyterLab is a popular notebook interface that is particularly suitable for this demonstration since it is implemented in Python. Once the JupyterLab task is started in a cluster, it automatically gets a public URL that the users can connect to. Cowait Notebooks allow data scientists to host notebook interfaces in any Kubernetes cluster with minimal setup. 
Executing Cowait Notebooks within a Cowait task lets the notebook access Cowait's underlying task scheduler and allow sub-tasks to be launched directly from the notebook cells -- data scientists can thus easily execute background tasks on the cluster. In the next section, we present Jakobsson and Henriksson's solution to allow access to local files -- and thus enabling work with local IDEs.

\subsection{Local Files and Cowait Notebooks Executing on Clusters}
Jakobsson and Henriksson developed a proof-of-concept implementation of a general solution to file sharing between a data scientist's local computer and software running on a remote cluster. The key enabler is a custom networked file system implemented using File System in Userspace (FUSE)\footnote{File System in Userspace, \url{https://github.com/libfuse/libfuse}}. FUSE is an interface for userspace programs to export a file system to the Linux kernel. To make the solution compatible with as many different data science applications as possible, the network file system was implemented as a custom storage driver for Kubernetes. Kubernetes is the most popular cluster orchestration solution, available as a managed service from all major cloud providers. Furthermore, Kubernetes is an open-source solution that users can also deploy on-premise. Practically, Jakobsson and Henriksson ensured compatibility with Kubernetes by implementing the Container Storage Interface, an open standard for developing new Kubernetes storage options\footnote{https://kubernetes-csi.github.io/docs/}.

The goal of the MSc thesis project was to design a user-friendly, reliable, and widely compatible solution to file sharing for data scientists. The aim was to provide seamless access to files residing on a data scientist's local computer for other data scientists accessing the local files through cloud-based notebook interfaces executing on Kubernetes clusters. With such a solution in place, data scientists could collaborate online using the notebook interfaces they prefer while allowing state-of-the-art software engineering tools to operate in IDEs on local machines.

To evaluate the proof-of-concept, Jakobsson and Henriksson conducted two separate studies. First, a quantitative study was carried out to verify the solution's performance in light of requirements set by prior user experience research on human response times ~\cite[p.~135]{humanResponseTimes}. The authors studied the performance as different numbers of files, of different sizes, where accessed under different network conditions. While details are available in the MSc thesis~\cite{backtick}, the general finding is that the solution satisfied the requirement of file access within 1 second for reasonable file sizes and realistic network latency. We consider this a necessary but not sufficient requirement for the novel solution.

Second, Jakobsson and Henriksson conducted a qualitative study to collect deep insights into the solution's utility. The authors recruited a mix of data scientists and software developers (with substantial ML experience) to perform a carefully designed programming task under a think-aloud protocol~\cite{thinkAloud}. The purpose was to collect feedback on whether the novel file sharing solution could improve the overall experience of working with cloud-based notebook interfaces. The feedback displayed mixed impressions. Data scientists who were comfortable using managed cloud solutions expressed hesitation to use such a system due to reduced ease-of-use and potential collaboration issues. The group that was the most positive were developers with a software engineering background, who were excited to be able to use familiar tooling for local files. Despite the mixed opinions, we still perceive the proof-of-concept as promising -- but more work is needed to bridge notebook interfaces and local IDEs. 

\section{MLOps -- A Key Enabler for Agility in Software 2.0} \label{sec:mlops}
Many organizations report challenges in turning an ML proof-of-concept into a production-quality AI system~\cite{bosch_engineering_2020}. The experimental nature of ML development limits qualities such as reproducibility, testability, traceability, and explainability –- which are needed when putting a trustworthy product or service on the market. On top of this, an AI system must be maintained until the product or service reaches its end-of-life. This holistic lifecycle perspective, i.e., what follows post-release, is often missing when novice data science teams develop AI proofs-of-concept in the sandbox. An organization must continuously monitor the ML models in operation and, in many cases, evolve the models according to feedback from the production environment -- where phenomena such as distributional shifts can be game-changers~\cite{sculley_hidden_2015}. Without designing for the operations phase and ensuring that ML model changes easily can be pushed to production, it will be tough to reach sustainably value-creating AI solutions. This attractive state is sometimes referred to as \textit{Operational AI}~\cite{tapia2018implementing}. In the next section, we will share our view on how the concept of MLOps can help organizations reach this state.

\subsection{Continuous Engineering in the AI Era}
In software development, continuous software engineering and DevOps emerged to reduce the lead time and remove the barriers between development, testing, and operations~\cite{Ebert2016}. Workflow automation in pipelines is fundamental, as it enables approaches such as 1) continuous integration (integration of code changes followed by test automation), 2) continuous delivery (building software for an internal test environment), and 3) continuous deployment (delivery of software to actual users)~\cite{fitzgerald_continuous_2017}. Depending on the application, organizations can also add staging processes when human validation is needed. Thanks to the automation, development qualities such as traceability come at a substantially lower cost compared to a manual workflow~\cite{jabbari2018towards}. DevOps has inspired a similar mindset within ML development in the form of \textit{MLOps}, i.e., the standardization and streamlining of ML lifecycle management~\cite{treveil2020introducing} -- which is a recommended approach to tackle continuous engineering in Software 2.0~\cite{hummer_modelops_2019}.

Just like DevOps is more than a set of tools, MLOps can be seen as a mindset on the highest level. As an engineering discipline, MLOps is a set of practices that combines ML, DevOps, and Data Engineering. Organizations adopting MLOps hope to deploy and maintain ML systems in production reliably and efficiently. Going beyond technology, MLOps involves embracing a culture with corresponding processes that an organization must adapt for the specific application domain. MLOps has emerged from the Big Tech Internet companies; thus, customization is required to fit smaller development organizations. Extrapolating from DevOps in conventional software engineering~\cite{Ebert2016,jabbari2018towards}, MLOps relies on pipeline automation to remove the barriers between data processing, model training, model testing, and model deployment.

MLOps is not yet well-researched from an academic perspective. The primary reason is that MLOps is a production concept, i.e., the phenomenon must be studied in the field rather than in university labs. However, this does not mean that MLOps should not be targeted by academic research. On the contrary, it is critically important that software and systems engineering researchers initiate industrial collaborations to allow empirical studies of what works and what does not when developing and evolving AI systems. As always in software engineering research, we have to identify the most important variation points needed to provide accurate guidance given specific application contexts. Just like there are uncountably many ways to implement pipeline automation -- the ML tools market is booming -- there is not a one-size-fits-all way to adopt MLOps in an organization.

\subsection{Reinforced AI Systems using Buttresses and Rebars}
Just as agile development enters regulated domains~\cite{diebold2018agile}, Software 2.0 is gradually entering critical applications~\cite{borg_safely_2019}. Examples include automotive software~\cite{falcini2017deep} and software in healthcare~\cite{Jiang230}. From a quality assurance perspective, AI systems using ML constitute a paradigm shift compared to conventional software systems. A contemporary deep neural network might be composed of hundreds of millions of parameter weights -- such an artifact is neither applicable to code reviews nor standard code coverage testing. Development organizations have learned how to develop trustworthy code-based software systems through decades of software engineering experience. This collected experience has successfully been captured in different industry standards. Unfortunately, many best practices are less effective when developing AI systems. Bosch \textit{et al.} and others argue that software and systems engineering must evolve to enable efficient and effective development of trustworthy AI systems~\cite{bosch_engineering_2020}. One response to this call is that new standards are under development in various domains to complement existing alternatives for high-assurance systems~\cite{vidot2021certification}.

Due to the growing reliance on AI systems, the European Union (EU) AI strategy stresses the importance of \textit{Trustworthy AI}. EU defines such systems as lawful, ethical, and robust~\cite{high-level_expert_group_on_artificial_intelligence_ethics_2019}. Unfortunately, we know that existing software engineering approaches such as requirements traceability~\cite{borg2017traceability} and verification \& validation~\cite{borg_safely_2019} are less effective at demonstrating system trustworthiness when functionality depends on ML models. Due to its experimental nature, data science makes it hard to trace design decisions after-the-fact and the resulting ML models become less reproducible~\cite{notebookPainPoints}. Moreover, the internals of ML models are notoriously difficult to interpret~\cite{gilpin2018explaining}, and AI systems are difficult to test~\cite{zhang_machine_2020,riccio_testing_2020}.

Not only must developers of critical AI systems comply with emerging industry standards, but novel AI regulations are also expected in the EU. In April 2021, the European Commission proposed an ambitious \textit{Artificial Intelligence Act} (AIA)~\cite{aiact}. AIA is a new legal framework with dual ambitions for turning Europe into the global hub for trustworthy AI. First, AIA aims to guarantee the safety and fundamental rights of EU citizens when interacting with high-risk AI systems. Second, AIA seeks to strengthen AI innovation by providing legal stability and instilling public trust in the technology. Many voices have been raised about the proposed legislation, in which especially the broad definition of AI has been criticized. However, all signs point to increased regulation of AI in the EU, in line with the now established General Data Protection Regulation~\cite{gdpr} -- including substantial fines defined in relation to annual global turnover. 

ML is an increasingly important AI technology in the digitalization of society that receives substantial attention in the AIA. According to the proposal, any providers of high-risk solutions using ML must demonstrate AIA conformance to an independent national authority prior to deployment on the EU internal market. Demonstrating this compliance will be very costly -- and how to effectively (and efficiently!) do it remains an important open research question.

\begin{figure}
\label{fig:reinforce}
\includegraphics[width=\textwidth]{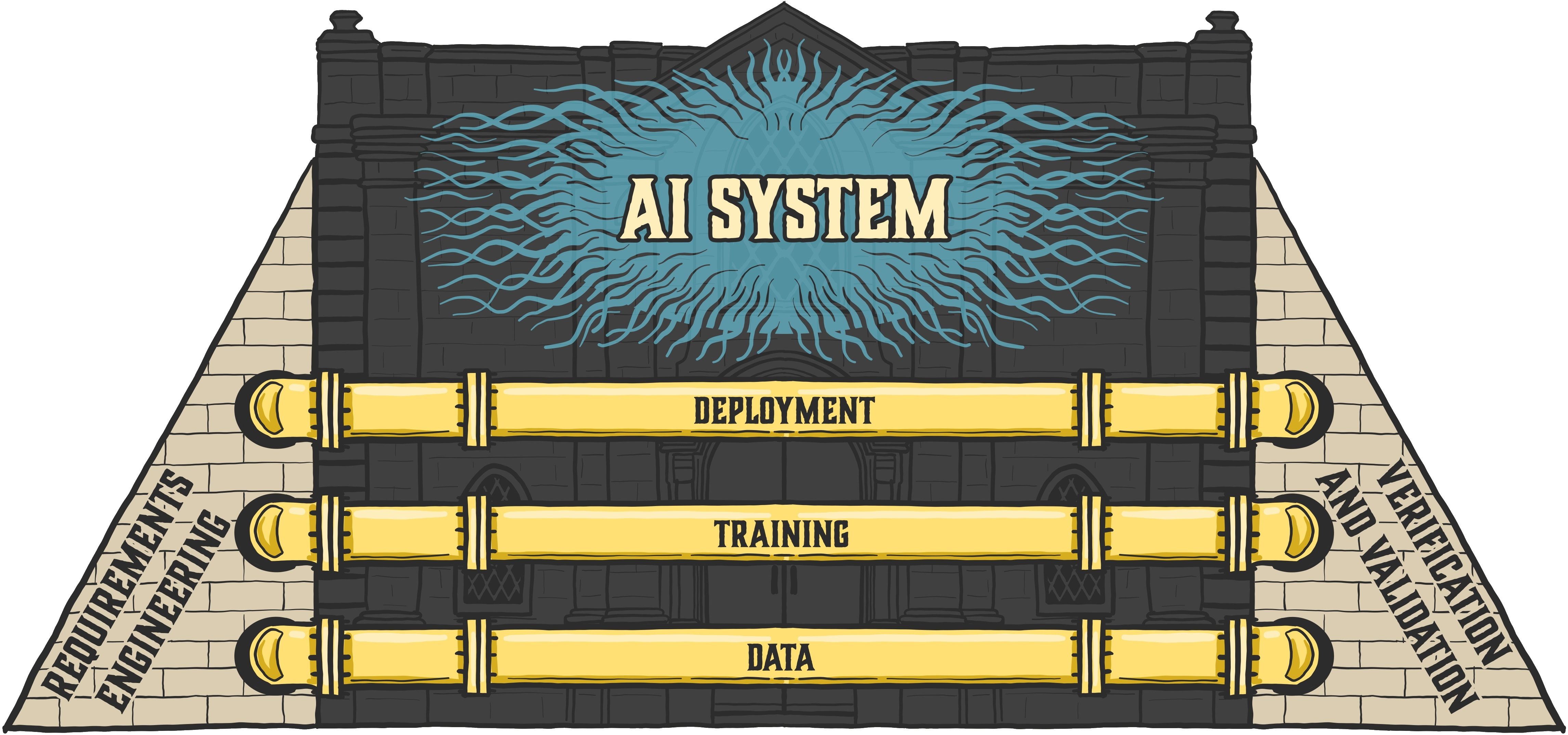}
\caption{Metaphorical buttresses and rebars. Robust requirements engineering and verification \& validation support the engineering of an ever-changing AI system. Pipeline automation in an MLOps context constitutes the rebars that sustain trustworthiness by strengthening the AI system despite the dynamics involved.}
\end{figure}

We are currently exploring the topic of built-in trustworthiness through a metaphor of reinforced engineering: \textit{buttresses and rebars}. Our fundamental position is that organizations must tackler quality assurance from two directions. Requirements engineering and verification \& validation shall work together like two bookends supporting the AI system, including its development and operations, from either end. Figure~\ref{fig:reinforce} illustrates how the primary reinforcement originates in buttressing the development of the AI system with requirements engineering (to the left) and verification \& validation (to the right). The metaphor, inspired by construction engineering, further borrows the concept of rebars, i.e., internal structures to strengthen and aid the AI system. In our metaphor, the rebars are realized in the form of so-called automation \textit{pipelines} for data, training, and deployment, respectively. Pipeline automation allows continuous engineering throughout the lifecycle, i.e., data management, training, deployment, and monitoring in an MLOps context. Pipeline automation enables flexibly adding automated quality assurance approaches as pipe segments, e.g., GradCAM heatmaps for explainability~\cite{borg2021test}, originating in the requirements engineering and verification \& validation buttresses. The envisioned reinforcement allows organizations to continuously steer the development and operations toward a trustworthy AI system –- in the context of highly agile data science, the CACE principle, and the ever-present risks of distributional shifts.

Numerous studies report that requirements engineering is the foundation of high-quality software systems. However, the academic community has only recently fully embraced the idea of tailored requirements engineering for AI systems. We argue that the particular characteristics of ML development in data science necessitate an evolution of requirements engineering processes and practices~\cite{vogelsang_requirements_2019}. New methods are needed when development transitions to the generation of rules based on training data and specific fitness functions. Based on a 2020 Dagstuhl seminar, K\"astner stressed requirements engineering as a particular ML challenge, Google researchers express it as underspecification~\cite{d2020underspecification}, and several papers have been recently published by the requirements engineering research community~\cite{ahmad2021s,habibullah2021non,siebert2021construction}. 

Academic research on verification \& validation tailored for AI systems has received a head start compared to requirements engineering for AI. New papers continuously appear, and secondary studies on AI testing~\cite{zhang2020machine,riccio2020testing} and AI verification~\cite{xiang2018verification} reveal hundreds of publications. As automation is close at hand for verification \& validation solutions, the primary purpose of the pipelines in the metaphor is to stress that they shall reach all the way to the requirements engineering buttress. Aligning requirements engineering with verification \& validation can have numerous benefits in software engineering~\cite{bjarnason2014challenges} -- and even more so, we argue, in AI engineering. Our planned next steps include exploring AIA conformant high-risk computer vision systems with industry partners reinforced by buttresses and rebars. Our ambition is to combine automated verification \& validation with an integrated requirements engineering approach~\cite{bjarnason2013integrated} in the continuous engineering of MLOps. Finally, we are considering introducing yet another metaphor from construction engineering, i.e., virtual plumblines as proposed by Cleland-Huang \textit{et al.} to maintain critical system quantities~\cite{cleland2008goal}. We posit that reinforcement and alignment will be two key essential concepts in future AI engineering, supported by a high level of automation to allow agile development of Software 2.0.

\section{Conclusion}
Whether we endorse the term Software 2.0 or not, AI engineering inevitably brings novel challenges. The experimental nature of how data scientists perform ML development means that the work must be agile. However, this agility can be supported in various ways. In this keynote address, we discussed two contemporary phenomena in data science and ML. First, we presented notebook interfaces, weaknesses, and a solution proposal to lower the bar for them to co-exist with modern IDEs. Second, we shared our perspective on MLOps and our ongoing work on providing reinforced engineering of AI systems in this context. Agility and continuous engineering are needed in AI engineering, as AI systems are ever-changing and often operate in dynamic environments. Finally, the EU AI Act further exacerbates the need for reinforced engineering and alignment between requirements engineering and verification \& validation. As a guiding light toward this goal, we introduced our vision of metaphorical buttresses and rebars.

\section*{Acknowledgements}
Martin Jakobsson and Johan Henriksson are the co-creators of the solution presented in Section~\ref{sec:notebooks} and deserve all credit for this work. Our thanks go to Backtick Technologies for hosting the MSc thesis project and Dr. Niklas Fors, Dept. of Computer Science, Lund University for acting as the examiner. This initiative received financial support through the AIQ Meta-Testbed project funded by Kompetensfonden at Campus Helsingborg, Lund University, Sweden and two internal RISE initiatives, i.e., ``SODA - Software \& Data Intensive Applications'' and ``MLOps by RISE.'' 

%
%
%
%
\bibliographystyle{splncs}
\bibliography{lasd2022}

\begin{thebibliography}{10}

\bibitem{giray2021software}
Giray, G.:
\newblock A software engineering perspective on engineering machine learning
  systems: State of the art and challenges.
\newblock Journal of Systems and Software \textbf{180} (2021)  111031

\bibitem{Ebert2016}
{Ebert}, C., {Gallardo}, G., {Hernantes}, J., {Serrano}, N.:
\newblock {DevOps}.
\newblock IEEE Software \textbf{33}(3) (2016)  94--100

\bibitem{diebold2018agile}
Diebold, P., Theobald, S.:
\newblock How is agile development currently being used in regulated embedded
  domains?
\newblock Journal of Software: Evolution and Process \textbf{30}(8)  e1935

\bibitem{hanssen2018safescrum}
Hanssen, G.K., St{\aa}lhane, T., Myklebust, T.:
\newblock SafeScrum{\textregistered}-Agile Development of Safety-Critical
  Software.
\newblock Springer Nature, Cham, Switzerland (2018)

\bibitem{kim2016emerging}
Kim, M., Zimmermann, T., DeLine, R., Begel, A.:
\newblock The emerging role of data scientists on software development teams.
\newblock In: Proc. of the 38th International Conference on Software
  Engineering. (2016)  96--107

\bibitem{sletholt2011we}
Sletholt, M.T., Hannay, J.E., Pfahl, D., Langtangen, H.P.:
\newblock What do we know about scientific software development's agile
  practices?
\newblock Computing in Science \& Engineering \textbf{14}(2) (2011)  24--37

\bibitem{sculley_hidden_2015}
Sculley, D.,  et~al.:
\newblock Hidden {Technical} {Debt} in {Machine} {Learning} {Systems}.
\newblock In: Proc. of the 28th International Conference on {Neural}
  {Information} {Processing} {Systems}. (2015)  2503--2511

\bibitem{amershi2019software}
Amershi, S., Begel, A., Bird, C., DeLine, R., Gall, H., Kamar, E., Nagappan,
  N., Nushi, B., Zimmermann, T.:
\newblock Software engineering for machine learning: A case study.
\newblock In: Proc. of the 41st International Conference on Software
  Engineering. (2019)  291--300

\bibitem{wan2019does}
Wan, Z., Xia, X., Lo, D., Murphy, G.C.:
\newblock How does machine learning change software development practices?
\newblock IEEE Transactions on Software Engineering \textbf{47}(9) (2021)
  1857--1871

\bibitem{vogelsang_requirements_2019}
Vogelsang, A., Borg, M.:
\newblock Requirements {Engineering} for {Machine} {Learning}: {Perspectives}
  from {Data} {Scientists}.
\newblock In: Proc. of the 27th {International} {Requirements} {Engineering}
  {Conference} {Workshops}. (2019)  245--251

\bibitem{notebookPainPoints}
Chattopadhyay, S., Prasad, I., Henley, A.Z., Sarma, A., Barik, T.:
\newblock What’s wrong with computational notebooks? {P}ain points, needs,
  and design opportunities.
\newblock In: Human Factors in Computing Systems. (2020)  1--12

\bibitem{backtick}
Jakobsson, M., Henriksson, J.:
\newblock Sharing local files with {K}ubernetes clusters (2021) MSc Thesis,
  Lund University,
  http://lup.lub.lu.se/student-papers/record/9066685/file/9066686.pdf.

\bibitem{notebookKnuth}
Knuth, D.E.:
\newblock Literate programming.
\newblock The Computer Journal \textbf{27}(2) (1984)  97--111

\bibitem{knuth_book}
Knuth, D.E.:
\newblock Literate Programming.
\newblock Center for the Study of Language and Information, Stanford, US (1992)

\bibitem{hannay2009scientists}
Hannay, J.E., MacLeod, C., Singer, J., Langtangen, H.P., Pfahl, D., Wilson, G.:
\newblock How do scientists develop and use scientific software?
\newblock In: Proc. of the ICSE Workshop on Software Engineering for
  Computational Science and Engineering, Ieee (2009)  1--8

\bibitem{notebookLandscape}
{Vognstrup Fog}, B., {Nylandsted Klokmose}, C.:
\newblock Mapping the landscape of literate computing.
\newblock In: Proc. of the 30th Annual Workshop of the Psychology of
  Programming Interest Group. (2019)

\bibitem{notebookForaging}
Kery, M.B., John, B.E., O'Flaherty, P., Horvath, A., Myers, B.A.:
\newblock Towards effective foraging by data scientists to find past analysis
  choices.
\newblock In: Human Factors in Computing Systems. (2019)  1--13

\bibitem{notebookUntangle}
Kery, M.B., Myers, B.A.:
\newblock Interactions for untangling messy history in a computational
  notebook.
\newblock In: Proc. of the IEEE Symposium on Visual Languages and Human-Centric
  Computing. (2018)  147--155

\bibitem{notebookManageMesses}
Head, A., Hohman, F., Barik, T., Drucker, S.M., DeLine, R.:
\newblock Managing messes in computational notebooks.
\newblock In: Human Factors in Computing Systems. (2019)  1--12

\bibitem{notebookNotes}
Singer, J.:
\newblock Notes on notebooks: {I}s {Jupyter} the bringer of jollity?
\newblock Proc. of the ACM SIGPLAN International Symposium on New Ideas, New
  Paradigms, and Reflections on Programming and Software (2020)  180--186

\bibitem{humanResponseTimes}
Nielsen, J.:
\newblock Usability Engineering.
\newblock Morgan Kaufmann Publishers, Burlington, MA, USA (1993)

\bibitem{thinkAloud}
Kuusela, H., Paul, P.:
\newblock A comparison of concurrent and retrospective verbal protocol
  analysis.
\newblock The American Journal of Psychology \textbf{113}(3) (2000)  387--404

\bibitem{bosch_engineering_2020}
Bosch, J., Holmstr{\"o}m~Olsson, H., Crnkovic, I.:
\newblock Engineering {AI} systems: {A} research agenda.
\newblock In: Artificial Intelligence Paradigms for Smart Cyber-Physical
  Systems.
\newblock IGI Global (2021)  1--19

\bibitem{tapia2018implementing}
Tapia, P., Palacios, E., No{\"e}l, L.,  et~al.:
\newblock Implementing {O}perational {AI} in telecom environments.
\newblock Tupl White Paper \textbf{7} (2018)

\bibitem{fitzgerald_continuous_2017}
Fitzgerald, B., Stol, K.J.:
\newblock Continuous software engineering: {A} roadmap and agenda.
\newblock Journal of Systems and Software \textbf{123} (2017)  176--189

\bibitem{jabbari2018towards}
Jabbari, R., Ali, N., Petersen, K., Tanveer, B.:
\newblock Towards a benefits dependency network for {DevOps} based on a
  systematic literature review.
\newblock Journal of Software: Evolution and Process \textbf{30}(11) (2018)
  e1957

\bibitem{treveil2020introducing}
Treveil, M., Omont, N., Stenac, C., Lefevre, K., Phan, D., Zentici, J.,
  Lavoillotte, A., Miyazaki, M., Heidmann, L.:
\newblock Introducing MLOps.
\newblock O'Reilly Media, Inc., Sebastopol, CA, USA (2020)

\bibitem{hummer_modelops_2019}
Hummer, W., Muthusamy, V., Rausch, T., Dube, P., Maghraoui, K.E., Murthi, A.,
  Oum, P.:
\newblock {ModelOps}: {Cloud}-{Based} {Lifecycle} {Management} for {Reliable}
  and {Trusted} {AI}.
\newblock In: Proc. of the {International} {Conference} on {Cloud}
  {Engineering}. (2019)  113--120

\bibitem{borg_safely_2019}
Borg, M., Englund, C., Wnuk, K., Duran, B., Levandowski, C., Gao, S., Tan, Y.,
  Kaijser, H., L\"onn, H., T\"ornqvist, J.:
\newblock Safely {Entering} the {Deep}: {A} {Review} of {Verification} and
  {Validation} for {Machine} {Learning} and a {Challenge} {Elicitation} in the
  {Automotive} {Industry}.
\newblock Journal of Automotive Software Engineering \textbf{1}(1) (2019)
  1--19

\bibitem{falcini2017deep}
Falcini, F., Lami, G., Costanza, A.M.:
\newblock Deep learning in automotive software.
\newblock IEEE Software \textbf{34}(3) (2017)  56--63

\bibitem{Jiang230}
Jiang, F.,  et~al.:
\newblock {Artificial Intelligence in Healthcare: Past, Present and Future}.
\newblock Stroke and Vascular Neurology \textbf{2}(4) (2017)  230--243

\bibitem{vidot2021certification}
Vidot, G., Gabreau, C., Ober, I., Ober, I.:
\newblock Certification of embedded systems based on machine learning: A
  survey.
\newblock arXiv preprint arXiv:2106.07221 (2021)

\bibitem{high-level_expert_group_on_artificial_intelligence_ethics_2019}
{High-Level Expert Group on Artificial Intelligence}:
\newblock Ethics {Guidelines} for {Trustworthy} {Artificial} {Intelligence}.
\newblock Technical report, European Commission, Brussels, Belgium (2019)

\bibitem{borg2017traceability}
Borg, M., Englund, C., Duran, B.:
\newblock Traceability and deep learning-safety-critical systems with traces
  ending in deep neural networks.
\newblock Proc. of the Grand Challenges of Traceability: The Next Ten Years
  (2017)  48--49

\bibitem{gilpin2018explaining}
Gilpin, L.H., Bau, D., Yuan, B.Z., Bajwa, A., Specter, M., Kagal, L.:
\newblock Explaining explanations: An overview of interpretability of machine
  learning.
\newblock In: Proc. of the 5th International Conference on Data Science and
  Advanced Analytics. (2018)  80--89

\bibitem{aiact}
{European Commission}:
\newblock Proposal for a {R}egulation of the {E}uropean {P}arliament and of the
  {C}ouncil laying down harmonised rules on artificial intelligence
  ({A}rtificial {I}ntelligence {A}ct) and amending certain union legislative
  acts (2021-04-21)
  \newline\url{https://eur-lex.europa.eu/legal-content/EN/TXT/?uri=CELEX%3A52021PC0206}.

\bibitem{gdpr}
{European Commission}:
\newblock Regulation ({EU}) 2016/679 of the {E}uropean {P}arliament and of the
  {C}ouncil on the protection of natural persons with regard to the processing
  of personal data and on the free movement of such data, and repealing
  {D}irective 95/46/ec ({General Data Protection Regulation}).
\newblock Official Journal of the European Union \textbf{119} (2016-05-04)
  1--88

\bibitem{borg2021test}
Borg, M., Jabangwe, R., {\AA}berg, S., Ekblom, A., Hedlund, L., Lidfeldt, A.:
\newblock Test automation with {Grad-CAM} heatmaps - {A} future pipe segment in
  {MLOps} for vision {AI}?
\newblock In: Proc. of the 14th International Conference on Software Testing,
  Verification and Validation Workshops. (2021)  175--181

\bibitem{d2020underspecification}
D'Amour, A., Heller, K., Moldovan, D., Adlam, B., Alipanahi, B., Beutel, A.,
  Chen, C., Deaton, J., Eisenstein, J., Hoffman, M.D.,  et~al.:
\newblock Underspecification presents challenges for credibility in modern
  machine learning.
\newblock arXiv preprint arXiv:2011.03395 (2020)

\bibitem{ahmad2021s}
Ahmad, K., Bano, M., Abdelrazek, M., Arora, C., Grundy, J.:
\newblock What’s up with requirements engineering for artificial intelligence
  systems?
\newblock In: Proc. of the 29th International Requirements Engineering
  Conference. (2021)  1--12

\bibitem{habibullah2021non}
Habibullah, K.M., Horkoff, J.:
\newblock Non-functional requirements for machine learning: Understanding
  current use and challenges in industry.
\newblock In: Proc. of the 29th International Requirements Engineering
  Conference. (2021)  13--23

\bibitem{siebert2021construction}
Siebert, J., Joeckel, L., Heidrich, J., Trendowicz, A., Nakamichi, K., Ohashi,
  K., Namba, I., Yamamoto, R., Aoyama, M.:
\newblock Construction of a quality model for machine learning systems.
\newblock Software Quality Journal (2021)  1--29

\bibitem{zhang2020machine}
Zhang, J.M., Harman, M., Ma, L., Liu, Y.:
\newblock Machine learning testing: Survey, landscapes and horizons.
\newblock IEEE Transactions on Software Engineering (2020)

\bibitem{riccio2020testing}
Riccio, V., Jahangirova, G., Stocco, A., Humbatova, N., Weiss, M., Tonella, P.:
\newblock Testing machine learning based systems: {A} systematic mapping.
\newblock Empirical Software Engineering \textbf{25}(6) (2020)  5193--5254

\bibitem{xiang2018verification}
Xiang, W., Musau, P., Wild, A.A., Lopez, D.M., Hamilton, N., Yang, X.,
  Rosenfeld, J., Johnson, T.T.:
\newblock Verification for machine learning, autonomy, and neural networks
  survey.
\newblock arXiv preprint arXiv:1810.01989 (2018)

\bibitem{bjarnason2014challenges}
Bjarnason, E., Runeson, P., Borg, M., Unterkalmsteiner, M., Engstr{\"o}m, E.,
  Regnell, B., Sabaliauskaite, G., Loconsole, A., Gorschek, T., Feldt, R.:
\newblock Challenges and practices in aligning requirements with verification
  and validation: {A} case study of six companies.
\newblock Empirical software engineering \textbf{19}(6) (2014)  1809--1855

\bibitem{bjarnason2013integrated}
Bjarnason, E.:
\newblock Integrated Requirements Engineering - Understanding and Bridging Gaps
  in Software Development.
\newblock Lund University, Sweden,
  https://lucris.lub.lu.se/ws/portalfiles/portal/3427902/4117182.pdf (2013)

\bibitem{cleland2008goal}
Cleland-Huang, J., Marrero, W., Berenbach, B.:
\newblock Goal-centric traceability: Using virtual plumblines to maintain
  critical systemic qualities.
\newblock IEEE Transactions on Software Engineering \textbf{34}(5) (2008)
  685--699

\end{thebibliography}
%

\end{document}